\documentclass[12pt]{article}
\usepackage{amssymb}
\usepackage{amstext}
\usepackage{fullpage}
\usepackage{epsfig}
\usepackage{graphicx}
\usepackage{colordvi}
\usepackage{hyperref}
\usepackage{url}
\usepackage{xspace}
\usepackage{amsmath}

\newcommand\ignore[2]{}

=165.mm

\usepackage{amssymb}
\usepackage{amsmath}
\usepackage{epsfig}
\usepackage{graphicx}
\usepackage{url}
\usepackage{xspace}

\newcommand{\ThirdNodeType}{\rput}
\usepackage{pstricks,pst-node,pst-tree}
\usepackage{wrapfig}
\psset{nodesep=3pt}
\psset{xunit=.9cm}
\psset{yunit=.9cm}
\begin{document}

\setlength{\parskip}{.2in}

\bibliographystyle{alpha}

\Large\begin{center} \textbf{Effectively Nonblocking Consensus
Procedures Can Execute Forever -- a Constructive Version of FLP
}\\
\vspace{.25in}
 \large{Robert L. Constable}\\
 \normalsize\em{Cornell University} \\
 \normalsize\em{Print date: \today}\\
 \normalsize\em{July 17, 2008}\footnote{The current draft includes improvements up to August 30,
2008 and small non technical edits to the July 17, 2008 version made in August 2011.  See Acknowledgements at the end of the article for grants supporting this work.}

\end{center}

\begin{flushleft}
\normalsize

\begin{abstract}
\noindent The Fischer-Lynch-Paterson theorem (FLP) says that it is
impossible for processes in an \emph{asynchronous distributed
system} to achieve consensus on a binary value when a single process
can fail; it is a widely cited theoretical result about network
computing. All proofs that I know depend essentially on classical
(nonconstructive) logic, although they use the hypothetical
construction of a nonterminating execution as a main lemma.
\flushleft
FLP is also a guide for protocol designers, and in that role there
is a connection to an important property of consensus procedures,
namely that they should not \emph{block}, i.e. reach a global state
in which no process can decide.
\flushleft
A deterministic fault-tolerant consensus protocol is
\emph{effectively nonblocking} if from any reachable \emph{global
state} we can find an execution path that decides. In this article
we effectively construct a nonterminating execution of any such
protocol. That is, given any effectively nonblocking protocol \textbf{P} and a natural
number $n$, we show how to compute the $n$-th step of an infinitely
\emph{indecisive computation} of \textbf{P}. From this fully constructive
result, the \emph{classical FLP} follows as a corollary as well as a
stronger classical result, called here \emph{Strong FLP}. Moreover,
the construction focuses attention on the important role of
nonblocking in protocol design.
\flushleft
An interesting consequence of the constructive proof is that we can, in principle, build an \emph{undefeatable attacker} for a consensus protocol that is provably correct, indeed because it is provably correct. We can do this in practice on certain kinds of networks.

\end{abstract}

\section{Introduction}
\subsection{Background}

The standard version of the Fisher-Lynch-Paterson theorem is that
there is no asynchronous distributed algorithm that is responsive to
its inputs, solves the agreement problem, and guarantees 1-failure
termination. This is a negative statement, producing a
contradiction, yet implicit in all proofs is an imagined
construction of a nonterminating execution in which no process
decides, they "waffle" endlessly. That imagined execution is an
interesting object, displaying what can go wrong in trying to reach
consensus and characterizing a class of protocols. The hypothetical
execution is used to guide thinking about consensus protocol design
(illustrated below). In light of that use, a natural question about
the classical proofs of FLP is whether the hypothetical infinite
waffling execution could actually be constructed from any purported
consensus protocol \textbf{P}, that is, given \textbf{P}, can we
exhibit an algorithm $\alpha$ such that for any natural number $n$,
$\alpha(n)$ is the $n$-th step of the indecisive computation.

It appears that no such explicit construction could be carried out
following the method of the classical proof because there isn't
enough information given with the protocol, and the key concept in
the standard proofs, the notion of \emph{valence} (\emph{univalence}
and \emph{bivalence}), is not defined effectively, i.e. they require
knowing the results of all possible executions. This means that the
case analysis used to imagine the infinite execution can not
actually be decided. Of course, it is not possible to find this
infinite execution by simply running a purported protocol. Only a
\emph{proof} can show that it will run forever.

Other authors \cite{Vol04,BW87} have reformulated the proof of the
FLP in a way that singles out the infinite computation as the result
of a separate lemma, but they do not provide an effective means of
building the infinite computation and do not use constructive
reasoning. I refer to Volzer's classical result as \emph{Strong
FLP}; it is a corollary of the effective construction given here.

The key to being able to build the nonterminating execution is to
provide more information, which we do by introducing the notion of
\emph{effective nonblocking}, defining bivalence effectively, and
introducing the idea of a \emph{v-possible execution}. We use the
term \emph{bivalence} in most of this article to make comparison
with the classical ideas clear, but when contrasting this work to
others, we will use the term \emph{effective bivalence}.
\footnote{In the original FLP article the authors say: Let C be a
configuration in an execution of the protocol, and let V be the set
of all decision values reachable from C. C is bivalent if V is
$\{0,1\}$ and univalent if V is $\{v\}$ for $v$ a Boolean.}

Effective nonblocking is a natural concept in the setting in which
we verify protocols using constructive logic, say the rules of the
Nuprl formal programming environment or of the Coq prover
\cite{BYC04}. The logic of Nuprl is Computational Type Theory (CTT)
\cite{ABCEKLM05}, which is constructive, and the logic of Coq is the
Calculus of Inductive Constructions (CIC), closely related to CTT
and also constructive. So when we prove that a protocol is
nonblocking, we obtain the effective witness function used in the
definition below. Mark Bickford in his Nuprl formalization
of consensus protocols has done formal
proofs of nonblocking from which Nuprl can extract the deciding
state and could extract an execution as well.

The importance of nonblocking can be seen from this "blocking
theorem" by Robbert van Renesse in \cite{Rvr08}: \emph{A consensus
protocol that guarantees a decision in the absence of failures may
block in the presence of even a single failure.} This is justified
by citing FLP, and it follows cleanly from CFLP as I show below. Robbert van Renesse says: ``Blocked states occur when one or more
processes fail at a time at which other processes cannot determine
if the protocol has decided. A protocol that tolerates failures must
avoid such blocked states''. Protocol designers actually carry out
an analysis of blocking in debugging designs. A constructive proof
of the \emph{blocking theorem} could find the blocking scenario
after designating a process that fails. Knowing precisely the number
of blocking scenarios and their properties would be useful in
evaluating protocol designs.

It is fascinating that once we use the concept of effective
bivalence, it is possible to automatically translate some
nonconstructive proofs of FLP into fully constructive ones from
which it is possible to build the nonterminating execution.  Here we look
at the simpler result that we can effectively build nonterminating
executions. These are executions that endlessly waffle about the
decisions that are possible, decisions actually taken by
\emph{decisive executions}.

Since it is not possible to provide an \emph{algorithm}, i.e. a
\emph{terminating} consensus procedure, we start with the kind of
protocol that can be built, and stress the possibility of
nontermination by calling it a \emph{procedure} not an algorithm.

\subsection{Computing Model}

The results here depend on the computing model behind the Logic of
Events, \cite{BC08} which is essentially embedding the standard model of asynchronous
message-passing network computing into
Computational Type Theory. The standard model is presented in the book
\emph{Distributed Computing} of Attyia \& Welch \cite{AW04} and is
similar to the one in \cite{FLP85}. We assume reliable FIFO communication
channels.

A \emph{global state} of the system consists of the state of the
processes and the condition of the message queues. An
\emph{execution} is an alternating sequence of global states and
actions taken by processes. Thus an execution $\alpha$ of
distributed system \textbf{S} determines sequence of \emph{global
states}, $s_1, s_2, s_3, ...$. These are also called
\emph{configurations} of the execution. Execution is fair \emph{in} that all messages sent to nonfailing processes
will eventually be read and all enabled actions will eventually be
taken by processes that do not fail.

A \emph{step of computation} can involve any finite number of processes
reading a message from an input channel, changing the internal
state, and sending messages on output channels.  In the proofs here,
we pick an order on these steps so that there is always a single
action separating the global states. We say that a \emph{schedule}
determines the order of the actions.

\subsection{Definitions}

\textbf{Definition}: A Boolean \emph{consensus procedure}  on
processes $P_{i}~i=1,...,n$ tolerating $t$ failures is a possibly
nonterminating distributed procedure \textbf{P} which is deterministic
(no randomness), responsive on uniform initializations, consistent
(all deciding processes agree on the same value).

\textbf{P} is called \emph{effectively nonblocking} if from any
reachable global state $s$ of an execution of \textbf{P} and any
subset $Q$ of $n-t$ nonfailed processes, \emph{we can find} an
execution $\alpha$ from $s$ using $Q$ and a process $P_{\alpha}$ in
$Q$ which decides a value $v~\epsilon~\mathbb{B}$.

Constructively this means that we have a computable function,
$wt(s,Q)$ which produces an execution $\alpha$ and a state
$s_\alpha$ in which a process, say $P_\alpha$ decides a value $v$.

In this setting, a consensus procedure is \emph{responsive} if when
\emph{all} processes are initialized to $v$, they terminate with
decision $v$ unless they fail.  This means that all nonblocking
witnesses will return $v$ as well.

The nonblocking property requires that consensus procedures
tolerating $t$ failures can use any subset of $n-t$ processes to
pick out from any partial execution a process that makes a decision.
This is enough information for an \emph{algorithmic adversary} to
prevent a deterministic consensus procedure, one that does not rely
on randomness, from terminating on every execution. The adversary
can keep adjusting the schedule of executions to prevent processes
from deciding.

It is important to have good notations for the class of all
processes of \textbf{P} except for $P_i$. We denote that class by $Q_i$, because we
want to factor executions into steps of a specified process and
those of the remaining processes. These are disjoint sets, and we
can combine executions from them by appending one to another and
infer joint properties from the separate properties of each.

\textbf{Definition:} For a $v~\epsilon~\mathbb{B}$, a global state
$s$ is \emph{v-possible} \emph{iff} for some subset $Q$ of $n-t$
processes we can find using the nonblocking witness a state
$s^{\prime}_{Q}$ and a process $P_Q$ in $s^{\prime}_{Q}$  that
decides $v$. That is, $wt(s,Q)$ produces a computation ending in
$s^{\prime}_{Q}$.

\textbf{Definition:} A global state $b$ is \emph{bivalent} iff we
can find executions $\alpha_0$ and $\alpha_1$ from $b$ that decide
$0$ and $1$ respectively.  We can pick out the deciding process from
the execution.  A state is \emph{bivalent via $Q_i$} if neither
execution involves a step of process $P_i$. Note, if $b$ is
bivalent, we can effectively exhibit the executions $\alpha_0$ and
$\alpha_1$.

\textbf{Fact:} It is \emph{decidable} whether the global states of a
consensus procedure are $v-possible$.

Note, we can't decide bivalence.

\subsection{Summary of Results}

\textbf{Initialization Lemma:} For any effectively nonblocking
consensus procedure \textbf{P} with $n >1$, there is a bivalent
initial global state $b_{0}$.

\textbf{One Step Lemma:} Given any bivalent global state $b$ of an
effectively nonblocking consensus procedure \textbf{P}, and any
process $P_{i}$, we can find a extension $b^\prime$ of $b$ which is
bivalent via $Q_i$.

\textbf{Theorem (CFLP):} Given any deterministic effectively
nonblocking consensus procedure \textbf{P} with more than two
processes and tolerating a single failure, we can effectively
construct a nonterminating execution of it.

We also say that \textbf{P} can endlessly waffle. The proof is to
use the Initialization Lemma to find a bivalent starting state $b_0$
and then use the One Step Lemma to create an unbounded sequence of
bivalent states.

\textbf{Corollary (FLP):}  There is no single-failure responsive,
deterministic consensus algorithm (terminating consensus procedure)
on two or more processes.

\textbf{Corollary (Strong FLP)*:} Given any nonblocking
deterministic consensus procedure on two or more processes, it has a
nonterminating execution.

\textbf{Corollary (Blocking)*:} If all executions of consensus
procedure \textbf{P} terminate in a decision when no process fails,
then there is a global state on which \textbf{P} blocks when one
process fails.

The asterisk means that the results are not constructive, they use
classical logic. To stress that an existence claim is not
constructive, we sometimes say that an object such as an execution
is constructed \emph{using magic}; this means that our proof
requires nonconstructive logical rules in showing that the object
exists, rules such as the law of excluded middle or proof by
contradiction or Markov's principle, or the classical axiom of
choice, etc.

\subsection{Relationship to the Original FLP Proof}

Some of these results correspond closely to the lemmas used in the
Fischer, Lynch, Paterson paper \cite{FLP85}.  For example, our
Initialization Lemma is their Lemma 2, our One Step Lemma is close
to their Lemma 3, and the Commutativity Lemma used in the next
section is their Lemma 1. Our FLP Corollary is their Theorem 1.  In
the proof of Theorem 1, they structure the argument around an
unstated Lemma 0 which in their words is essentially ``...we
construct an admissible run that avoids ever taking a step that
would commit the system to a particular decision.'' They call these
runs \emph{forever indecisive}.

If they had defined a consensus procedure as above and had stated
nonblocking classically, this lemma would be:\emph{Any nonblocking
consensus procedure has forever indecisive executions}, which I call
Strong FLP; it is close to Volzer's classical result \cite{Vol04}.
Instead, Fischer, Lynch, and Paterson get nonblocking from assuming
at the start for the sake of contradiction the existence of a
terminating consensus algorithm. We can see the Strong FLP result
emerging by factoring out an assumption they need from assuming the
existence of a terminating protocol and packaging it into an
explicit statement of a ``Lemma 0''. I hope to discuss, in the future, this technique of
``refactoring'' theorems to make them constructive.

\section{Proofs}
\subsection{Key Lemmas}

\textbf{Fact:} It is \emph{decidable} whether the global states of a
consensus procedure are $v-possible$.

To decide whether a state is $v-possible$ we note that the
definition of effective nonblocking provides a function, say $wt$
that takes the state and a subset of $n-t$ processes and asks for
each such subset whether the deciding state decides $0$ or $1$.  It
is useful to introduce a notation for sets of processes that do not
include a particular process $P_{i}$; let $Q_{i}$ be all processes
of \textbf{P} except for $P_{i}$. Given state $s$, we make this
decision for processes tolerating one failure by computing
$wt(s,Q_{1})$,...,$wt(s,Q_{n})$.

\textbf{Initialization Lemma:} For any effectively nonblocking
consensus procedure \textbf{P}, there is a bivalent initial global
state $b_{0}$.

\textbf{Proof}

The argument for this is similar to the one used in the classical
FLP result, but we employ the decision of witnesses rather than a
purported consensus algorithm to find evidence for bivalence. We
first note that if all processes are initialized by $v$, then by
responsiveness, the consensus procedure must terminate with decision
$v$, and all nonblocking witnesses decide $v$. So if the initial
state is all $0$, then the witness decides $0$ and likewise for $1$.

Now consider a sequence of initial states where we start from the
all $0$ initialization, call it $s_0$ and progressively change the
initialization, processes by process, from $0$ to $1$ until we reach
the initialization of all $1$'s. Let these states be $s_0$, $s_1$,
..., $s_n$, where $n$ is the number of processes. For each initial
state, we ask whether there is a $1$ deciding state produced by the
witness function, which must happen by the time we reach the
initialization of all $1$'s.

Let $s_k$ be the first state where a decision is $1$, say
$wt(s_k,Q_m)$ decides $1$ for some $m$, and note that $k>0$, $P_k$
is initialized to $1$ for the first time, and the process $P_{k+1}$
is still initialized to $0$ if $k < n$ .

Consider the computation $\alpha$ from $wt(s_{k-1},Q_k)$ in which
process $P_k$ does not participate and the decision is $0$. We can
replay this from $s_k$. To the processes participating, this
computation will look like one with $P_k$ initialized to $0$, i.e.
one from $s_{k-1}$, and we have found an execution that results in a
$0$ decision from $s_k$ as we need to prove, that is $s_k$ is
bivalent. Take $b_0 = s_k$.

\textbf{Qed}

In the classical argument, one assumes that the procedure \textbf{P}
terminates, and on $s_k$ a computation $\alpha$ terminates with $1$
for the first time in the sequence. The next step is to alter the
schedule and produce a new computation $\alpha^\prime$ in which
$P_k$ is slow and does not affect the decision.  In this case the
computation looks just like one in which $P_k$ is initialized to
$0$, so the result is as for $s_{k-1}$, the value is $0$. Thus $s_k$
is bivalent.

The next lemma is the heart of the argument. We use it in the main
theorem, CFLP, to build a round-robin schedule in which each process
takes a step from one bivalent state to another, thus generating an
unbounded sequence of states in which no process decides. In
addition to the proof given below, I also include in the last
section of the article a program that shows the computational
content of this proof and also an elegant condensed version of the
proof that David Guaspari produced in response to this proof and its
algorithm.

\textbf{One Step Lemma:} Given any bivalent global state $b$ of an
effectively nonblocking consensus procedure \textbf{P}, and any
process $P_{i}$, we can find a extension $b^\prime$ of $b$ which is
bivalent via $Q_i$.

\textbf{Proof}

If we knew that bivalent $b$ was already bivalent via $Q_i$, we
would be done. First, we can calculate one deciding state using
$wt(b,Q_i)$; suppose that is $d_0$ which decides $0$ at the end of
execution $\alpha_0$. Since $b$ is bivalent, we also have an
execution $\alpha_1$ that decides $1$ and may take steps in process
$P_i$ (see figure \ref{lemma-A}).

Our plan now is to move backwards from $d_1$ along execution
$\alpha_1$ step by step toward state $b$ using the processes in
$\alpha_1$, which include process $P_i$, looking for a state
$b^\prime$ which is bivalent via $Q_i$ (see figure \ref{lemma-B}). We first find
a state and a computation such that the final steps to a $1$
decision don't involve any $P_i$ steps.

Suppose that the last step to $d_1$ is from state $u$ via $P_k$ for
$k \neq i$ by action $a$, then we have a $1$ decision using $Q_i$
from $u$ as we wished, and we will check to see if $wt(u,Q_i)$
computes a $0$ decision.  If so we are done. Otherwise we look at
the next process step in ${\alpha}_1$. Before we look at the method
of moving from $u$ back toward $b$, we need to consider how to
handle $P_i$ steps, so look at the case when the last step to $d_1$
was taken by $P_i$, i.e. $k = i$.

If $k = i$, then we look for a new path via $Q_i$ to a $1$ decision.
Compute $wt(u,Q_i)$ and let the deciding state be $d^\prime$ by
execution $\beta$ (see figure \ref{lemma-C}).  We claim that $d^\prime$ must
decide $1$. To see this, notice that by the Commutativity Lemma
below, $\beta$ followed by action $a$ of $P_i$ leads to the same
state as action $a$ followed by computation $\beta$, that is
$a{\beta}(u) = {\beta}a(u)$ (as in figure \ref{lemma-C}). But since $d_1$ is a
deciding state, $a{\beta}(u)$ must also decide $1$ by the Agreement
property of \textbf{P}. Then the execution ${\beta}a$ must decide
$1$ as well. So by Agreement applied to $d^\prime$, that deciding
state must decide $1$. Now $\beta$ is a $Q_i$ path that decides $1$,
and we have moved one step closer to $b$ on the path ${\alpha}_1$.

Now we keep moving back from $u$ along ${\alpha}_1$ toward $b$
showing that we can maintain a path via $Q_i$ to a state that
decides $1$ and looking for a $Q_i$ path to a $0$ deciding state. We
will find such a path, namely ${\alpha}_0$ by the time we reach $b$
if not before.

As we move back from $u$ toward $b$ on ${\alpha}_1$,  suppose we
encounter a $P_k$, step $k \neq i$ with action $a$, say going from
state $s$ to $s^\prime$. We know from the construction that
$wt(s^\prime,Q_i)$ does not lead to a $0$ decision, and we look at
the predecessor state $s$, and compute $wt(s,Q_i)$. If $0$ is
decided, and $k \neq i$, then we are done, and we take $b^{\prime} =
s$. However, if $k = i$ then we need a different analysis.

Thus suppose we find a state $s^\prime$ reached by an action $a$ of
$P_i$.  Notice that there is  by the construction so far a
computation from $s^\prime$ to a $1$ decision via $Q_i$, either
along some $\beta$ or along ${\alpha}_1$.

Now compute $wt(s, Q_i)$ and let the result be $d^\prime$, a
deciding state.  We consider two cases based on the decision at
$d^\prime$.

If $d^\prime$ decides $0$, then let $\alpha^{\prime}$ be the
computation from $s$ to $d^\prime$. We can use Commutativity and
Agreement to show that this computation can be replayed from
$s^\prime$ with same results, a $0$ decision. This is a witness that
$s^\prime$ is bivalent via $Q_i$ and finishes the construction, with
$b^\prime = s^\prime$ (see figure \ref{lemma-D}).

If $d^\prime$ decides $1$, then we have a new execution via $Q_i$,
say $\beta$ from $s$ to a $1$ deciding state, say $d_{1}^\prime$.
Moreover, we have taken another step closer to $b$ along $\alpha_1$.

We continue in this manner, incorporating $P_k$ steps into the
$\alpha_1$ path or building a new $\beta$ path to a $1$ deciding
state until we either reach $b$ or find a state $s$ before then that
is bivalent via $Q_i$.

\textbf{Qed}

Here are diagrams of the constructions we just described. In the
section on further details and alternatives, I also include a
program that executes the computation implicit in this proof.

\begin{figure}[h]
  \centering
\hspace{.1\linewidth}\begin{minipage}{0.4\linewidth}
\ThirdNodeType(0,-2){\rnode{d1}{$d_1$}}
\ThirdNodeType(2,0){\rnode{b}{$b$}}
\ThirdNodeType(4,-2){\rnode{d0}{$d_0$}}
\psset{nodesep=3pt}
\ncline{d1}{b}
\naput{$Q_j$}
\nbput{$\alpha_1$}
\ncline{b}{d0}
\naput{$Qi$}
\nbput{$\alpha_0$}
\vspace{1.4in} 
\caption{lemma-A}
                \label{lemma-A}
 \end{minipage}\hfill
 \begin{minipage}{0.5\linewidth}
\psset{nodesep=2pt}
\ThirdNodeType(0,-4){\rnode{d1}{$d_1$}}
\ThirdNodeType(2,-2){\cnode*{.1}{bP}}
\uput{.2cm}[90](2,-2){\texttt{$b'$}}
\ThirdNodeType(4,-4){\rnode{d0P}{$d'_0$}}
\ThirdNodeType(4,0){\rnode{b}{$b$}}
\ThirdNodeType(8,-4){\rnode{d0}{$d'_0$}}
\ncline{<-}{d1}{b}\naput[labelsep=.5cm]{$Q_j$}
\ncline{->}{bP}{d0P}
\ncline{->}{b}{d0}\naput{$Qi$}\nbput{$a_0$}
\vspace{1.4in} 
  \caption{lemma-B}
    \label{lemma-B}
 \end{minipage}
\end{figure}

\begin{figure}
 \begin{minipage}{0.5\linewidth}
\ThirdNodeType(0,-2){\rnode{d1}{$d_1$}}
\ThirdNodeType(2,0){\rnode{U}{$U$}}
\ThirdNodeType(4,-6){\rnode{dPP}{$d''$}}%
\ThirdNodeType(6,-4){\rnode{dP}{$d'$}} %
\ncline{<-}{d1}{U}\naput{$P_i$}\nbput{$a$}
\ncline{->}{U}{dP}\naput{$Q_i$}\nbput{$\beta$}
\ncline[linestyle=dashed]{->}{d1}{dPP}\nbput{$\beta$}
\ncline[linestyle=dashed]{<-}{dPP}{dP}\naput{$a$}\nbput{$P_i$}
\vspace{2.1in} 
  \smallskip
  \caption{lemma-C}
  \label{lemma-C}
\end{minipage}
\begin{minipage}{0.5\linewidth}
\ThirdNodeType(0,-4){\rnode{D}{${\tt decides\; 1}$}}
\ThirdNodeType(2,-2){\rnode{SP}{$S'$}}
\ThirdNodeType(4,0){\rnode{S}{$S$}}
\ThirdNodeType(6,-6){\rnode{dPP}{$d''$}}
\ThirdNodeType(8,-4){\rnode{dP}{$d'$}}
\ncline{->}{SP}{D}
\ncline{<-}{SP}{S}\naput{$P_i$}
\ncline{->}{S}{dP}\naput{$Q_i$}\nbput{$\alpha'$}
\ncline[linestyle=dashed]{->}{SP}{dPP}\nbput{$\alpha'$}
\ncline[linestyle=dashed]{<-}{dPP}{dP}
\vspace{2.1in} 
  \smallskip
  \caption{lemma-D}
  \label{lemma-D}
\end{minipage}
\end{figure}

\pagebreak

\subsection{Main Theorem (Constructive FLP) and Corollaries}

\textbf{Theorem (CFLP):} Given any deterministic effectively
nonblocking consensus procedure, we can find an infinite execution.

\textbf{Proof}

The unbounded execution $\alpha$ starts with a bivalent initial
state $b_0$ known to exist by the Initialization Lemma.  We now
schedule a round-robin execution of each process $P_i$ and action
$a$ extending the current bivalent state, say $s_k$, to a state
$b^\prime$ which is bivalent via $Q_i$ by the One Step Lemma.  At
this state, we apply the action $a$ of $P_i$ unless it has already
been applied in reaching $b^\prime$.  We can show that $m(b^\prime)$
is also bivalent via $Q_i$ by the Commutativity Lemma, and thus we
can repeat the construction using another process, say $P_j$ and its
enabled action. We compute $wt(m(b^\prime),Q_j)$ and look for a
witness with the opposite value, $wt(m(b^\prime),Q_m)$ or use the
$Q_i$ execution at $m(b^\prime)$ with the opposite valence.

Now find an extension that is bivalent via $Q_j$ using again the One
Step Lemma. In this manner we fairly execute steps of all processes,
yet never reach a deciding state.

\textbf{Qed}

\textbf{Corollary (FLP):}  There is no single-failure responsive
deterministic consensus algorithm (terminating consensus procedure).

\textbf{Proof}

Assume that \textbf{A} is such an algorithm. Let $b_0$ be a bivalent
initial state. Algorithm \textbf{A} is the nonblocking witness for
any reachable state, thus \textbf{A} is a consensus procedure, and
thus does not terminate.  So it is false that such an algorithm
exists according to the CFLP Theorem.

\textbf{Qed}

Note, this result is constructive, and its content is a
contradiction, not an infinite execution.

\textbf{Corollary:} If consensus procedure \textbf{P} is effectively
nonblocking, then we can find nonterminating executions even if no
process fails.

We note that in our construction of an infinite computation that
does not decide, none of the processes fails.

\textbf{Corollary (Strong FLP)*} If consensus procedure \textbf{P}
is nonblocking, then some execution of it is infinite.

We use the axiom of choice and the law of excluded middle to build a
noncomputable witness function for nonblocking and then follow the
construction in CFLP.

\textbf{Corollary (Blocking)*:} Given a consensus procedure
\textbf{A} that terminates when there are no failures, there is by
magic a computation that blocks (from which no decision is possible)
when a single process fails.

\textbf{Proof}

Because all executions of \textbf{A} must terminate when no process
fails, and because for nonblocking protocols there is always a
nonterminating execution even when no process fails, \textbf{A}
cannot be nonblocking.  Thus, by classical logic, there is a
blocking global state.

\textbf{Qed}

\subsection{Further Details and Alternatives}

There are other technical details and further intuitive insights
behind the lemmas that are worth presenting.

\paragraph{Initializations} The following notations help us make the
Initialization Lemma more compact. Let $s_j$ be the initialization
in which $P_i$ is initialized to $1$ for all $i \leq j$ and $P_i$ is
initialized to $0$ for all $i>j$ for $i=1,...,n$.

To find the first $s_k$ where $wt(s_k, Q)= 1$ for some $Q$, we
evaluate $wt(s_i, Q_j)$ systematically, increasing $i$ after trying
all subsets $Q_j$ for that $i$.  We know that these witnesses must
eventually produce a $1$ value because when $k=n$, then $wt(s_k, Q)=
1$ for all $Q$.

Let $s_k$ be the first initialization producing the decision $1$
using the nonblocking witness, say $wt(s_k,Q_m)$ decides one. Notice
that $wt(s_j, Q_i) = 0$ for all $j<k$ and all $i$ in $1 \leq i \leq
n$, and in particular, $wt(s_{k-1}, Q_k) = 0$, say by execution
$\alpha_0$. If for some $Q$ we have $wt(s_k,Q)$ decides $0$, then we
are done.  If not, we can replay computation $\alpha_0$ from $s_k$
in which process $P_k$ is scheduled to run very slow and not
participate in the decision. To the processes participating, this
computation will look like one with $P_k$ initialized to $0$, and
there will thus be an execution that results in a $0$ decision from
$s_k$ as we need to prove.

It seems natural to argue that $wt(s_{k-1}, Q_k) = wt(s_k, Q_k)$
since $P_k$ does not participate and the states differ only on $P_k$
initializations, but we do not impose conditions on the witness
about how it computes, so from $s_k$ the algorithm might produce a
different computation, say with a different schedule on the
participating processes.  However, we can replay the computation
from $s_{k-1}$ as in the above proofs.

\paragraph{Effective Bivalence}

In proving the One Step Lemma we need a key property of disjoint
sets of processes called commutativity.  It is this.

\textbf{Simple Commutativity Lemma:} Let $s$ be a global state and
consider disjoint sets of processes, ${P_i}$ and $Q_i$. Suppose
there is a computation $\alpha_1$ from $s$ using $Q_i$ to state
$s_1$ and computation $\alpha_2$ from $s$ using $P_i$ to state
$s_2$. Then there is a global state $s^\prime$ and a computation
from $s_1$ via $P_i$ to $s^\prime$ and from $s_2$ to $s^\prime$ via
$Q$.

\textbf{Proof}

We can think of $\alpha_2(\alpha_1(s)) = s^\prime =
\alpha_1(\alpha_2(s))$ because the two computations are disjoint and
can be ordered in either way, and we can delay messages from $P_i$
to the processes in $Q_i$ so that the two computations do not
interact.

\textbf{Qed}

\textbf{Commutativity Lemma:} Let $s$ be a global state and let $Q$
and $\bar{Q}$ be disjoint sets of processes. Suppose there is a
computation $\alpha_1$ from $s$ using $Q$ to state $s_1$ and
computation $\alpha_2$ from $s$ using $\bar{Q}$ to state $s_2$. Then
there is a global state $s^\prime$ and a computation from $s_1$ via
$\bar{Q}$ to $s^\prime$ and from $s_2$ to $s^\prime$ via $Q$.

This result follows by induction from the simple case by delaying
all messages between the disjoint sets, thus $\alpha_2(\alpha_1(s))
= s^\prime = \alpha_1(\alpha_2(s))$ because the two computations are
disjoint and can be ordered in either way.

\textbf{Alternative Proof of the One Step Lemma}

David Guaspari provided the following elegant compressed account of
the previous proof of the One Step Lemma. It reveals quite clearly
how simple the constructive proof of the FLP theorem can be, hence
how simply the FLP result can be explained. Its simplicity suggests
that it is worth applying the technique to open problems in
distributed computing and to simplifying known proofs.

By definition, a bivalent state $b$ can fork into different
execution paths to $0$ and $1$ decisions.  Call a pair of these
paths, say $(\alpha,\beta)$ a \emph{fork}. We call a fork an
\emph{i-fork} when one of the paths does not involve any steps of
process $P_i$ and a \emph{full i-fork} when neither path involves
steps of $P_i$.

The way we use forks in the One Step Lemma introduces an asymmetry
on the paths.  There will be a distinguished process $P_i$ for which
we are seeking a full $i$-fork.  For a bivalent state it is trivial
to find an $i$-fork for any $i$ by just computing $wt(b,Q_i)$ and
using that result as one branch.  To simplify managing this
asymmetry, we agree that the $\beta$ branch of an $i$-fork will be
the one without steps from $P_i$.  The $\alpha$ path may or may not
have $P_i$ steps.  If $\phi$ is an $i$-fork, let $i-len(\phi)$ be
the number of $P_i$ steps on the $\alpha$ path.  Then $\phi$ is a
full $i$-fork iff $i-len(\phi) = 0$.

\textbf{Fork Modification Lemma:} Let $\phi$ be an $i$-fork at state
$s$ with $i-len(\phi) = m > 0$. Suppose $a_m$ is the last $P_i$
action in the $\alpha$ branch, taking state $s_{m-1}$ to state
$s_m$. Let $v$ be the decision reached by $wt(s_{m-1}, Q_i)$, then:

1. If $v$ is the decision reached by $\beta$, we can effectively
construct a full $i$-fork from $s_m$, and

2. If $v$ is the decision reached by $\alpha$, we can effectively
construct an $i$-fork $\phi^\prime$ from $s$ such that
$i-len(\phi^\prime) < i-len(\phi)$.

\textbf{Proof}

For notational convenience, suppose that the $\beta$ path decides
$0$. Figure~\ref{i-fork} shows the $i$-fork $\phi = (\alpha,\beta)$,
together with $wt(s_{n-1},Q_i)$.  We have, in a slightly informal
notation:
\begin{itemize}
\item
$\alpha = \delta \cdot a_n \cdot \varepsilon$
\item
$\gamma$ is the sequence returned by $wt(s_{n-1},Q_i)$ and $b$ is
its final state
\item
$a_n$ is an action of process $P_i$
\item
none of the sequences $\beta$, $\gamma$, or $\varepsilon$ contains
an action from process $P_i$
\item
$d_1$ decides 1 and $d_0$ decides 0
\end{itemize}

\begin{figure}[h]
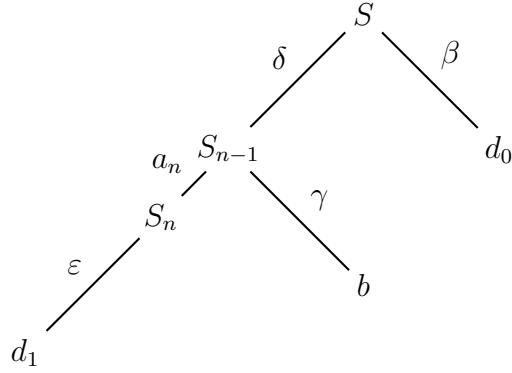

\hspace{2in}\begin{minipage}{0.5\linewidth}
\ThirdNodeType(0,-5){\rnode{d1}{$d_1$}}
\ThirdNodeType(5,0){\rnode{s}{$S$}}
\ThirdNodeType(3,-2){\rnode{S}{$S_{n-1}$}}
\ThirdNodeType(2,-3){\rnode{Sn}{$S_n$}}
\ThirdNodeType(7,-2){\rnode{d}{$d_0$}}
\ThirdNodeType(5,-4){\rnode{b}{$b$}}
\psset{nodesep=3pt}
\ncline{d1}{Sn}
\naput{$\varepsilon$}
\ncline{Sn}{S}
\naput{$a_n$}
\ncline{S}{s}
\naput{$\delta$}
\ncline{S}{b}
\naput{$\gamma$}
\ncline{s}{d}
\naput{$\beta$}
\end{minipage}
\vspace{1.7in} 
\caption{An $i$-fork\label{i-fork}}
\end{figure}

Case 1: In this case $b$ decides 0.  Consider
figure~\ref{commuting}. Because $a_n$ is an action of process $P_i$
and $\gamma$ contains no actions from $P_i$ the parallelogram
commutes, and the paths $a_n \cdot \gamma$ and $\gamma \cdot a_n$
lead to the same state, $c$, which must decide 0 because $b$ does.
So $(\varepsilon,\gamma)$ is a full $i$-fork from $s_n$.

\begin{figure}[h]
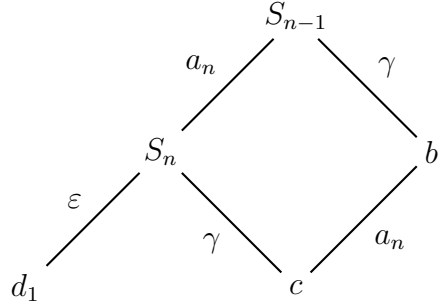

\hspace{2in}\begin{minipage}{0.5\linewidth}
\ThirdNodeType(0,-4){\rnode{d}{$d_1$}}
\ThirdNodeType(2,-2){\rnode{s}{$S_n$}}
\ThirdNodeType(4,0){\rnode{S}{$S_{n-1}$}}
\ThirdNodeType(4,-4){\rnode{c}{$c$}}
\ThirdNodeType(6,-2){\rnode{b}{$b$}}
\psset{nodesep=3pt}
\ncline{d}{s}
\naput{$\varepsilon$}
\ncline{s}{S}
\naput{$a_n$}
\ncline{s}{c}
\nbput{$\gamma$}
\ncline{S}{b}
\naput{$\gamma$}
\ncline{c}{b}
\nbput{$a_n$}
\end{minipage}
\vspace{1.7in} 
\caption{A commuting diagram\label{commuting}}
\end{figure}

Case 2: In this case, $b$ decides 1.  Then $\phi' =
(\delta\cdot\gamma, \beta)$ is an $i$-fork at $s$ and $i-len(\phi')
< i-len(\phi')$.

\textbf{Qed}

\pagebreak
\textbf{One Step Lemma}

Given any fork at $s$ and any $i$, we can effectively construct a
state $s'$ reachable from $s$ and a full $i$-fork at $s'$.

\textbf{Proof}

Let $(\alpha,\beta)$ be a fork at $s$ and let $\gamma$ be the
execution sequence returned by $wt(s,Q_i)$.  Then either
$(\alpha,\gamma)$ or $(\beta,\gamma)$ is an $i$-fork.  Now apply
Fork Modification repeatedly.

\textbf{Qed}

\textbf{A Program for the One Step Lemma}

The computational content of the One Step Lemma is a program whose
input is a bivalent state and a process $P_i$ and whose output is a
state that is bivalent via $Q_i$.

Logical Conditions: b is bivalent; $\alpha_1$ is an execution path
to $d_1$; $\alpha_0$ is an execution path to $d_0$; $P_i$ is the
designated process; $P_k$ is any process.

Program Variables and Code Segments:

\begin{itemize}
\item $S$, $S'$ denotes global states on path $\alpha_1$ from $b$ to
$d_1$.
\item $P$ is the process taking $S$ to $S'$
\item $Path$ is the execution path from $S'$ to a state deciding $1$
\item $pred(P)$ finds the predecessor process on $\alpha_1$
\item $pred(S)$ finds the predecessor state on $\alpha_1$, e.g.
$pred(S') = S$.
\item $Advance$ is the code $P := pred(P); S':= S; S := pred(S)$
(This code finds the next step moving toward $b$ on $\alpha_1$.)
\end{itemize}

Invariants:

\begin{itemize}
\item I0. $pred(S') = S$
\item I1. $Path$ is a $Q_i$ path from $S'$ to a $1$ deciding state.
\item I2. There is no $Q_i$ path known yet from $S'$ to a $0$
deciding state.
\item I3. Initially $S$ is $d_1$.
\end{itemize}

\begin{figure}
\begin{tabbing}
{\bf Begin} \= (Move along $\alpha$, from $d$, toward $b$)\\
\> While\= $(S \neq b\; \& \;wt(S,Q_i)$ finds execution path $\beta$ to decide(1)) {\bf do}\\
\> \> {\bf decide} \= [$P \stackrel{?}{=}P_i;$\\
\> \> \>{\bf case} $P = P_i (S \stackrel{P_i}{\longrightarrow} S')$ {\bf then} $Path := \beta ; Advance;$\\
\> \> \>{\bf case} $P = P_k (k \neq i) (S \stackrel{P_k}{\longrightarrow} S')$ {\bf then} $Path : = k; Advance]$\\
\>{\bf od}\\
\>{\bf if} $s = b$ {\bf then stop} ($b' = b$, $\alpha_0$ is path to $d_0$ deciding 0, Path decides ($i$))and is in $Q_i$\\
\>{\bf if} $wt(S,Q_i)$ decides 0 by path $\alpha'$ {\bf then} \\
\> \>{\bf decide} $(P \stackrel{?}{=}P_i$; )\\
\> \> \> {\bf case} $P = P_i$ {\bf then stop} \= {\bf b}\textbf{$'$} = {\bf S}\textbf{$'$}, $\alpha'$ is path to decide (0) from $S'$\\
\> \> \> \>by commutativity argument to carry\\
\> \> \> \> $\alpha'$ to state $S'$, Path is a $Q_i$ path to decide (1)\\
\> \> \>{\bf case} $P = P_k$ then $Path := Path \; P_k$; {\bf stop}\= \, \=({\bf b}\textbf{$'$} = {\bf s}, $\alpha'$ is\\
\> \> \> \> \> path to decide(0))\\
\> \> \> \> \>Path is path to decide(1))\\
{\bf End}
\end{tabbing}
\caption{One Step Program\label{One Step Program}}
\end{figure}

\pagebreak
\textbf{Acknowledgements}

This work was funded by NSF grant CNS 872612445 and by the
Information Directorate of the Air Force Research Lab (AFRL) at
Rome, grant FA 8750-08-2-0153.

I want to thank Robbert VanRenesse for explaining consensus
algorithms to me, for drawing my attention to the importance of the
nonblocking property, and for studying my argument -- a help in
making it more succinct. I also want to thank Mark Bickford for
responding to my early ideas for this proof and Uri Abraham for
listening to my argument and sharing his course notes on the FLP
theorem and providing pointers to the literature.  Shlomi Dolev
pointed out possible practical uses of the results that I continue
to examine. David Guaspari was very helpful in reading my proof and
drawing attention to points that needed clarification or correction.
He also provided a very elegant condensation of the One Step Lemma
that I sketched in the last section.

\end{flushleft}


\end{document}